# HEAT AND MASS TRANSFER IN THE POROUS WICK OF A CAPILLARY EVAPORATOR

Laetitia Mottet[1], Typhaine Coquard[2]
and Marc Prat

[1]laetitia.mottet@imft.fr
Université de Toulouse, INPT, UPS, IMFT, Avenue Camille Soula, 31400, Toulouse, France,
and CNRS, IMFT 31400, Toulouse, France

[2]Airbus Defence and Space, France, Avenue des Cosmonautes, 31402, Toulouse, France

## ABSTRACT

Heat and mass transfer inside the porous wick of a capillary evaporator is studied using a mixed pore-network model. The impact of the thermal conductivity of the wick on the overheating limit (defined as the difference between the maximum temperature at the top of the metallic casing and the saturated temperature), breakthrough (which occurs when the vapor reaches the wick inlet) and the parasitic heat flux lost by conduction at the entrance of the wick (which decreases the efficiency of the evaporator) is investigated. The study suggests a bilayer wick as a possible better design to optimize the performance of the evaporator. With this design, the inlet layer is of low thermal conductivity with small pore size so as to reduce the parasitic heat flux. The inlet layer also plays a role of capillary lock limiting the risk of breakthrough. The second layer, right under the metallic casing, is more conductive with a high thermal conductivity and larger pores so as to .limit the risk of overheating. It is shown that this design increases the range of heat loads which can be applied to the evaporator.

## INTRODUCTION

Loop heat pipes (LHP) ([8]) are increasingly used for cooling electronics devices in many sectors (aerospace, aeronautical, automobile…). As shown in Figure 1, a LHP consists of a capillary evaporator, a vapor line, a condenser, a liquid line and a compensation chamber. This study focuses on the evaporator. Figure 2 shows a cross-section of a cylindrical evaporator. During the operation of a LHP, a heat flux is applied on the external surface of the metallic casing. This induces the vaporization of the liquid and the formation of menisci at the surface or within the porous wick. The resulting capillary action sets the fluid in motion in the system. As sketched in Fig.2, the wick is fed in liquid by the liquid core linked to the compensation chamber (reservoir in Fig.2).

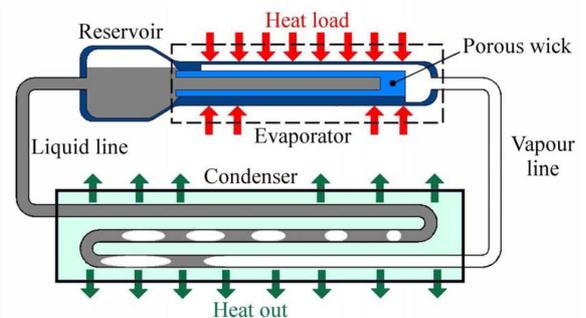

Figure 1: Schematic of a typical loop heat pipe (LHP)

The vapor produced inside the evaporator is evacuated thanks to vapor grooves.

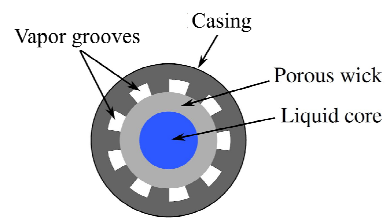

Figure 2: Cross section of a cylindrical capillary evaporator

The study of capillary evaporators has motivated many studies. Cao and Faghri [1] proposed an analytic solution to determine the temperature and the pressure field inside the porous wick when the wick is fully liquid saturated. Kaya and Goldak [6] developed a numerical study on heat and mass transfer within the wick. They notably discussed the boiling limit. Li and Peterson [7] developed 3D simulations also for a fully liquid saturated wick. Cao and Faghri [2] proposed a pseudo three dimensional model where the liquid flow inside the wick is computed as a 2D problem and the vapor flow inside the grooves is considered as a fully 3D problem. After a discussion on the boiling limit, they concluded that a 2D model is satisfying as long as the vapor velocity is sufficiently low. In addition to a continuum model with a sharp liquid-vapor interface, Figus et al. [5] developed a pore network model for studying the transfers in the wick. Coquard [3] developed a simplified pore network model, referred to as mixed pore network model because macroscopic parameters such as the permeability or the effective thermal conductivity were directly used as input parameters. A similar model is used in the present paper. The model enables one to obtain steady-state solutions.

## NOMENCLATURE

| | | |
|---|---|---|
| $c_p$ | = | Specific heat capacity |
| $h$ | = | Convection heat transfer coefficient |
| $k$ | = | Thermal conductivity |
| $K$ | = | Permeability |
| $L$ | = | Latent heat |
| $M$ | = | Molar mass |
| **n** | = | Unit normal vector |
| $P$ | = | Pressure |
| $Q$ | = | Heat flux |
| $R$ | = | Universal gas constant |
| $r$ | = | Pore radius |
| $T$ | = | Temperature |
| **u** | = | Velocity |

*Greek Symbols*

| | | |
|---|---|---|
| $\rho$ | = | Density |
| $\mu$ | = | Viscosity |
| $\sigma$ | = | Interfacial tension |

*Subscripts*

| | | |
|---|---|---|
| c | = | Casing |
| g | = | Groove |
| l | = | Liquid |
| p | = | Parasitic |
| v | = | Vapor |
| w | = | Wick |
| cap | = | Capillary |
| conv | = | Convection |
| evap | = | Evaporation |
| max | = | Maximum |
| sat | = | Saturation state |
| sub | = | Subcooling |

## 1 Mathematical model

The mathematical formulation is essentially the same as in Coquard [3]. The geometry of the computational domain is represented in Figure 3. It approximately corresponds to the domain delimited by the red dashed line in Figure 2. As can be seen in Figure 3, the vapor grooves are inside the casing. A discretized form of Darcy's law combined with the continuity equation is used for computing the vapor and liquid flows inside the porous wick neglecting gravity effects. Note that the flow in the vapor grooves is not explicitly computed. The thermal problem is solved both inside the wick and the casing.

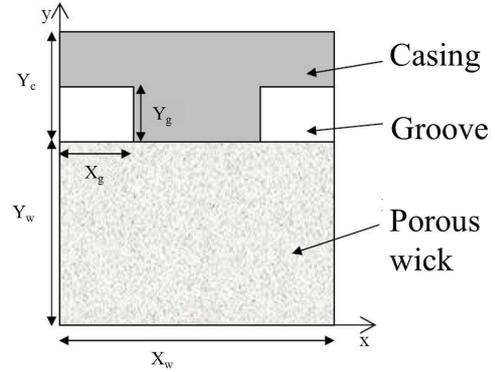

Figure 3: Computational domain

The various equations can be listed as follows.
Vapor region inside the wick:

$$\nabla . \rho_v \mathbf{u_v} = 0 \quad (1)$$

$$\mathbf{u_v} = -\frac{K}{\mu_v} \nabla P \quad (2)$$

$$\rho_v = \frac{PM}{RT} \quad (3)$$

$$(\rho c_p)_v \mathbf{u_v} . \nabla T = \nabla . (k_{w,v} \nabla T) \quad (4)$$

Liquid region inside the wick:

$$\nabla . \mathbf{u_l} = 0 \quad (5)$$

$$\mathbf{u_l} = -\frac{K}{\mu_l} \nabla P \quad (6)$$

$$(\rho c_p)_l \mathbf{u_l} . \nabla T = \nabla . (k_{w,l} \nabla T) \quad (7)$$

Heat transfer occurs only by conduction inside the casing.

$$\nabla . (k_c \nabla T) = 0 \quad (8)$$

The Clausius-Clapeyron relationship is used to determine the liquid-vapor interface temperature at saturation pressure.

$$P_{sat}(T_{sat}) = P_{ref} \exp\left(-\frac{LM}{R}\left[\frac{1}{T_{sat}} - \frac{1}{T_{ref}}\right]\right) \quad (9)$$

As the liquid-vapor interface can be inside the wick and/or at the wick/groove location, two different boundaries conditions have to be expressed. The boundary conditions at the liquid-vapor interface inside the wick are expressed as

$$T_l = T_v = T_{sat}(P_v) \quad (10)$$
$$\rho_l \mathbf{u_l}.\mathbf{n} = \rho_v \mathbf{u_v}.\mathbf{n} \quad (11)$$
$$(-k_{w,l}\nabla T).\mathbf{n} = (-k_{w,v}\nabla T).\mathbf{n} + L\rho_l \mathbf{u_l}.\mathbf{n} \quad (12)$$

At the wick/groove interface, the boundary condition reads

$$(-k_{w,l}\nabla T).\mathbf{n} = L\rho_l \mathbf{u_l}.\mathbf{n} \quad (13)$$

when the wick is liquid saturated.
A convective boundary condition is implemented at the wick/groove location when the wick is in vapor saturated,

$$(k_{w,v}\nabla T).\mathbf{n} = h_{w,g}(T - T_g) \quad (14)$$

where the convection heat transfer coefficient $h_{w,g}$ is calculated using a relationship given by Sleicher and Rouse [9].
The boundary condition at the wick-casing interface is defined by a zero normal velocity condition and the thermal flux continuity condition,

$$\nabla P.n = 0 \quad (15)$$
$$(-k_{w,i}\nabla T).\mathbf{n} = (-k_c \nabla T).\mathbf{n}, \text{ with i=l or v} \quad (16)$$

A convective boundary condition is applied at the casing/groove interface.

$$(k_c \nabla T).\mathbf{n} = h_{c,g}(T - T_g) \quad (17)$$

where $h_{c,g}$ is calculated in the same way as $h_{w,g}$.
The inlet pressure and temperature of the liquid at the entrance of the wick are imposed,

$$T = T_{sat} - \Delta T_{sub}, \quad P = P_{sat}(T_{sat}) \quad (18)$$

A heat load is applied at the external surface of the metallic casing,

$$(k_c \nabla T).\mathbf{n} = Q \quad (19)$$

Spatially periodic boundary conditions are applied on lateral sides of computational domain.
Note that $k_{w,i}$ with i=l or v represents the effective thermal conductivity of the wick defined as a function of the porosity, the solid phase thermal conductivity and the thermal conductivity of phase i.

## 2 Numerical method
A mixed pore network method is used as in Coquard [3]. The pore network is a two-dimensional square network of pores interconnected by throats. The equations are discretized using a finite volume like method adapted to the pore network geometry. The energy equation, eq.(4), is solved using a standard upwind scheme for the convective term. A home-made code written in Fortran90 was developed to solve the entire problem using an iterative-method. When the heat load applied on the casing is sufficiently low, the wick remains fully saturated by the liquid and the phase-change occurs at the groove – wick interface. When the temperature difference between the liquid under the fin and the saturation temperature is greater than a specified nucleation temperature, vapor forms within the porous structure and the wick becomes unsaturated. When this happens, it is assumed that the vapor quickly invades the first row of pores right under the casing (as first proposed by Demidov and al. [4]). From this initial position of the liquid-vapor interface within the wick, the governing equations are solved and the pressure jump $\Delta P = P_v - P_l$ along the liquid-vapor interface is computed. For each meniscus along the interface, the corresponding value is compared to the capillary pressure threshold $\Delta P_{cap} = 2\sigma/r$, which represents the maximum pressure jump compatible with a stable meniscus at the entrance of the considered throat. When the condition $\Delta P < \Delta P_{cap}$ is satisfied at each meniscus, the two-phase distribution is stable and the calculation stops. Otherwise ($\Delta P > \Delta P_{cap}$), a new pore is invaded and the procedure is repeated until convergence, i.e. until all menisci along the interface are mechanically stable.

## 3 Results
The working fluid is ammonia ($T_{sat} = 283K$). The following properties were used unless otherwise mentioned: $k_c = 22W/m/K$, $K = 4.10^{-14}m^2$, $\varepsilon = 0.25$.
The throat sizes were randomly distributed in the range [2-5 μm]. The nucleation temperature marking the onset of vaporization in the wick was equal to $T_{sat}$ + 3K. No subcooling was considered at the entrance ($\Delta T_{sub} = 0K$). Geometrical dimensions (as shown in Figure 3) are $X_w = 10mm$, $X_g = 2mm$, $Y_g = 1,5mm$, $Y_w = 5mm$ and $Y_c = 3mm$. The grid dimension is 107x200 points (21400 computational nodes).

### 3.1 Thermal conductivity and parasitic flux
The parasitic heat flux $Q_p$ is defined as the heat flux lost by conduction at the entrance of the wick. A desirable objective is to make this parasitic flux as low as possible in order to maximize the efficiency of the capillary evaporator. Figure 4 shows the evolution of the parasitic flux as a function of the heat load for different wick conductivities computed thanks to our pore network model.
As can be seen, the parasitic flux is low, less than 1% of the applied flux, for low wick thermal conductivity ($k_w = 0,2W/m/K$). By contrast, the parasitic flux is much higher and reaches the range of 10%-20% for a conductive wick ($k_w = 60 W/m/K$). Thus, these results suggest that a wick of low thermal conductivity is desirable to reduce the parasitic flux.

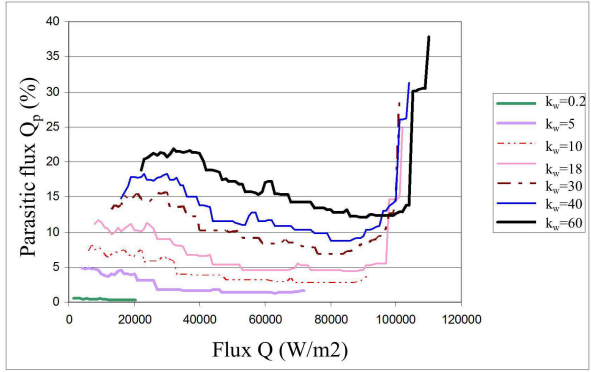

Figure 4: Parasitic heat flux as a function of the heat load applied on the casing for different wick thermal conductivities.

### 3.2 Thermal conductivity and casing overheating

The maximum temperature of this system is at the top right and left corners of the casing in Figure 3. This temperature should not exceed some specified value so as not to damage the cooled electronic device. The quantity $\Delta T_{max} = T_{max} - T_{sat}$ is defined as the overheating temperature. This is the difference between the maximum temperature and the saturated temperature (which corresponds to the groove temperature). Figure 5 shows the variation of $\Delta T_{max}$ as a function of the heat load for different wick thermal conductivities.

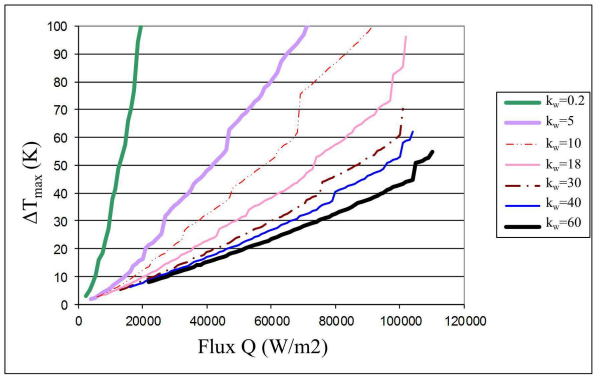

Figure 5: Casing overheating as a function of the applied heat load for different wick thermal conductivities.

For an insulating wick ($k_w = 0,2 W/m/K$), the overheating limit is reached quickly, whereas the overheating limit is not the operating limit for a conductive wick ($k_w = 60 W/m/K$), vapor breakthrough is more likely as discussed in § 3.3. In this respect, a conductive wick sounds a better choice.

### 3.3 Thermal conductivity and vapor saturation

Another important parameter is the vapor saturation inside the porous wick. Here, the saturation is defined as the ratio between the number of pores occupied by the vapor and the total number of pores. Its variation is represented in Figure 6. Vapor breakthrough occurs when the saturation is greater than about 85%,

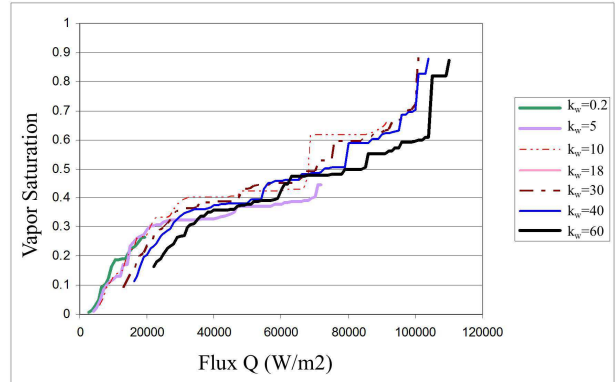

Figure 6: Evolution of the vapor saturation as a function of the applied heat load for different wick thermal conductivities.

For an insulating wick, the overheating limit is expected to be reached before the vapor breakthrough. By contrast, breakthrough occurs before reaching the overheating limit for a sufficiently conductive wick. Also the comparison between Figures 4 and 6 indicates that the abrupt increase in the parasitic heat flux occurs when the breakthrough is reached.

### 3.4 Bilayer wick

As shown in previous sections, the casing overheating and the parasitic flux are two important limiting parameters. Insulating wick ensures a weak heat loss by conduction at the entrance of the wick whereas a conductive wick reduces the risk of overheating. This suggests considering a bilayer wick combining a conductive layer and an insulating layer. As an example, consider the bilayer wick sketched in Figure 7.

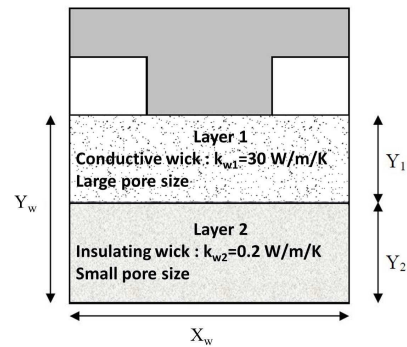

Figure 7: Schematic model of a bilayer wick

The first layer has a high thermal conductivity ($k_{w1} = 30 W/m/K$) to limit the overheating of the casing and relatively large pore sizes ($4 \pm 2 \mu m$) with a permeability $K_1 = 5,22.10^{-14} m^2$. The second layer has a low thermal conductivity ($k_{w2} = 0,2 W/m/K$) so as to limit the parasitic heat loss and small pores ($1.5 \pm 0.5 \mu m$). The permeability of the second layer is $K_2 = 7,34.10^{-15} m^2$. The quantity $e = Y_1/Y_w$ (see Figure 7) is used to characterize the size of each layer. Figure 8 shows that the layer 2 behaves as a capillary lock delaying the breakthrough.

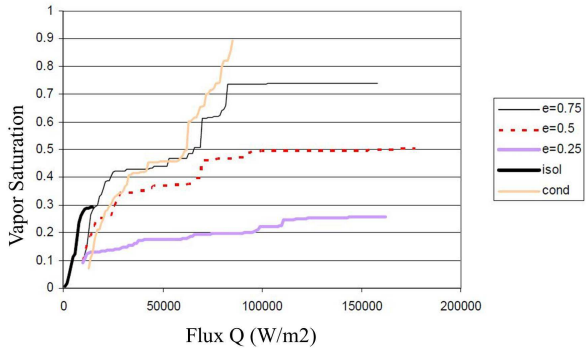

Figure 8: Evolution of the vapor saturation as a function of the heat load applied on the casing for different thicknesses of layer; "isol" curve and "cond" curve correspond to a single layer wick with the properties of the layer 2 and layer 1 respectively.

It is expected that the vapor saturation decreases with the size of the first layer. Indeed, the second layer also plays the role of a capillary lock which is well evidenced in Figure 8. The overheating limit is reached much later with the bilayer design as shown in Figure 9. The high conductivity of layer 1 has a clear positive effect. It seems that the overheating of the casing is even lower than when the entire wick is conductive. However, the size of the first conductive layer in the bilayer arrangement does not seem to have a significant incidence on $\Delta T_{max}$.

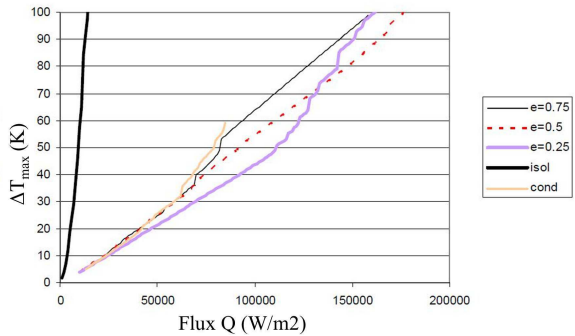

Figure 9: Variation of casing overheating as a function of the applied heat load for different thicknesses of layers.

As shown in Figure 10, the parasitic heat flux is greatly decreased with the bilayer wick compared to the fully conductive single layer wick (the flux lost by conduction for a fully conductive wick is about 10-40% as shown in Figure 4).

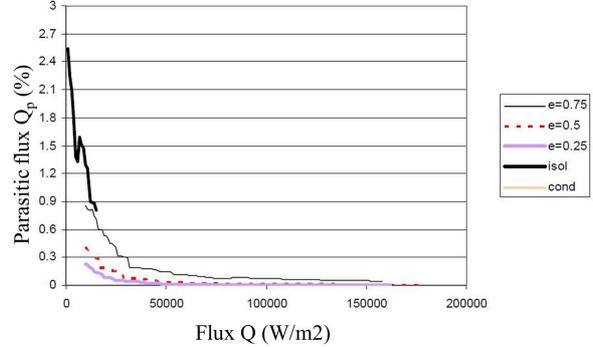

Figure 10: Variation of parasitic flux as a function of applied heat load on the casing for different thicknesses of layers.

The flux lost by conduction is less than 1% independently of the layer thicknesses with the bilayer wick. As shown in Figure 10, the greater the thickness of the insulating layer is, the lower the parasitic heat flux is.

## 4 Discussions

All the simulations presented in the previous section were performed using a two-dimensional square network. Although we believe that this allows obtaining correct qualitative results, it is well know that the structure of the liquid-vapor zone can be fundamentally different between 2D networks and 3D networks because of the different percolation properties. Thus, it is important to develop 3D pore network simulations.

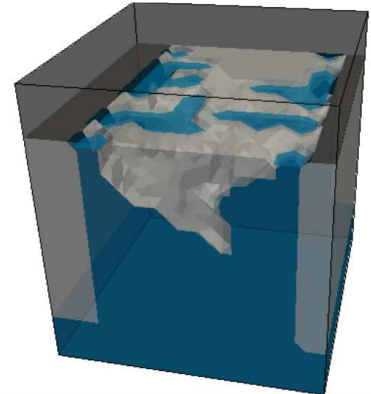

Figure 11: Example of phase distribution within the wick obtained using a 3D cubic pore network. Liquid in blue; vapor in grey.

Figure 11 shows an example of phase distribution obtained in a 3D cubic network. Note the coexistence of liquid and vapor regions right under the metallic casing in contrast with the 2D simulations, e.g. [5]. The results obtained with the 3D simulations will be presented and discussed in a forthcoming paper.

## CONCLUSIONS

Using two-dimensional pore network simulations, we presented a parametric study on the impact of the evaporator wick thermal conductivity on key parameters such as the casing overheating or the parasitic heat flux. The study suggests that a bilayer porous wick should increase the efficiency of the capillary evaporator. The layer right under the fin of the metallic casing has to be conductive so as to limit the overheating of the casing. The second layer, adjacent to the wick inlet, should be insulating with relatively smaller pores so as to prevent the risk of vapor breakthrough and to limit the parasitic heat flux lost by conduction. According to our simulations, the two layer wick would enable to reach the operating limits of the system for greater heat loads compare to single layer wicks.

However, it must be pointed out that the 2D nature of simulations can be questioned since it is likely that it favors the formation of a dry vapor zone adjacent to the casing fin within the wick compared to 3D pore network simulations. This issue is currently under study with the development of 3D simulations.

## ACKNOWLEDGEMENT

Financial supports from CNES and Astrium are gratefully acknowledged.